\documentclass{ifae_mod}
\usepackage{axodraw}

\newcommand{\be}{\begin{equation}}
\newcommand{\ee}{\end{equation}}
\newcommand{\br}{\begin{eqnarray}}
\newcommand{\er}{\end{eqnarray}}
\newcommand{\ba}{\begin{array}}
\newcommand{\ea}{\end{array}}
\newcommand{\bi}{\begin{itemize}}
\newcommand{\ei}{\end{itemize}}
\newcommand{\bn}{\begin{enumerate}}
\newcommand{\en}{\end{enumerate}}
\newcommand{\bc}{\begin{center}}
\newcommand{\ec}{\end{center}}

\def\MSbar{\overline{\mathrm{MS}}}

\def\pr#1 #2 #3 { {\rm Phys. Rev.}            {\bf #1}   (#2) #3}
\def\prd#1 #2 #3{ {\rm Phys. Rev.}            {\bf D#1}  (#2) #3}
\def\prl#1 #2 #3{ {\rm Phys. Rev. Lett.}      {\bf #1}   (#2) #3}
\def\plb#1 #2 #3{ {\rm Phys. Lett.}           {\bf B#1}  (#2) #3}
\def\npb#1 #2 #3{ {\rm Nucl. Phys.}           {\bf B#1}  (#2) #3}
\def\prp#1 #2 #3{ {\rm Phys. Rep.}            {\bf #1}   (#2) #3}
\def\zpc#1 #2 #3{ {\rm Z. Phys.}              {\bf C#1}  (#2) #3}
\def\epjc#1 #2 #3{ {\rm Eur. Phys. J.}        {\bf C#1}  (#2) #3}
\def\mpl#1 #2 #3{ {\rm Mod. Phys. Lett.}      {\bf A#1}  (#2) #3}
\def\ijmp#1 #2 #3{{\rm Int. J. Mod. Phys.}    {\bf A#1}  (#2) #3}
\def\ptp#1 #2 #3{ {\rm Prog. Theor. Phys.}    {\bf #1}   (#2) #3}
\def\jhep#1 #2 #3{ {\rm J. High Energy Phys.} {\bf #1}   (#2) #3}
\def\jphg#1 #2 #3{ {\rm J. Phys.}             {\bf G#1}  (#2) #3}
\def\cpc#1 #2 #3{ {\rm Comput. Phys. Commun.} {\bf #1}   (#2) #3}
\def\hepph#1{ [{\tt hep-ph/#1}]}

\def\as{\alpha_{\mathrm{S}}}
\def\aem{\alpha_{\mathrm{EM}}}
\def\ycut{y_{\mathrm{cut}}}
\def\lsim{\:\raisebox{-0.5ex}{$\stackrel{\textstyle<}{\sim}$}\:}
\def\gsim{\:\raisebox{-0.5ex}{$\stackrel{\textstyle>}{\sim}$}\:}
\def\thrust{\mbox{T}}
\def\Thrust{\mathrm{\tiny T}}

\begin{document}

%--- Header (title, authors, abstract)
\begin{header}

  % The title of your talk
  \title{Weak Corrections to Hadronic Observables}

  % List of authors with affiliations
  \begin{Authlist}
        E.~Maina\Iref{to},
        S. Moretti\Iref{southampton},
        M.R. Nolten\Iref{southampton},
        D.A. Ross\Iref{southampton}.

  \Affiliation{to}{Universit\`a di Torino and INFN, Torino, Italy}
%  \Affiliation{cern}{CERN, Geneva, Switzerland}
  \Affiliation{southampton}{University of Southampton, Southampton, UK}
%  \Acknowfoot{a}{On leave from University of Somewhere}
  \end{Authlist}

  % Collaboration (if applicable)
%  \collaboration {On Behalf of the XYZ Experiment}
  
  % Abstract
  \begin{abstract}
We illustrate one-loop weak corrections  to three-jet production
in $e^+e^-$ at $\sqrt{s}=M_Z$, to the production of a $Z$ or $\gamma$ in
association with a hard jet at hadron colliders and to the production cross
section of two $b$-jets at Tevatron and Large Hadron Collider
(LHC). 
  \end{abstract} 
  
\end{header}

% Beging of the text (no page break)
\section{Introduction}
One-loop EW corrections, as
compared to the QCD ones, have a relatively large impact. This can be understood 
(see Refs.~\cite{Melles:2001ye}--\cite{Denner:2001mn} and references therein
for reviews) in terms of the so-called
Sudakov (leading) logarithms of the form 
$\alpha_{\mathrm{W}}\log^2(\sqrt{\hat{s}}/M_{W}^2)$, which appear
in the presence of higher order weak corrections
(hereafter, $\alpha_{\rm{W}}\equiv\alpha_{\rm{\small EM}}/\sin^2\theta_{\rm W}$,
with $\alpha_{\rm{\small EM}}$ the Electro-Magnetic (EM) coupling constant and
$\theta_{\rm W}$ the weak mixing angle).
These `double logs' are due to a lack of cancellation of infrared (both soft
and collinear) virtual and real emission in
higher order contributions due to $W$-exchange
in spontaneously broken non-Abelian theories.

The problem is, in principle, present also in QCD. In practice, however, 
it has no observable consequences, because of the averaging on the 
colour degrees of freedom of partons, forced by their confinement
into colourless hadrons. This does not occur in the EW case,
where, e.g., the initial state can have a non-Abelian charge,
dictated by the given collider beam configuration. Modulo the
effects of the Parton Distribution Functions (PDFs), which spoil the subtle
cancellations among subprocesses with opposite non-Abelian charge,
 for example, this argument holds
for an initial quark doublet in proton-(anti)proton scatterings. These
logarithmic corrections (unless the EW process is mass-suppressed)
are universal (i.e., process independent) and are finite (unlike in
QCD), as the masses of the EW gauge bosons provide a physical
cut-off for $W$-boson emission. Hence, for typical experimental
resolutions, softly and collinearly emitted weak bosons need not be included
in the production cross-section and one can restrict oneself to the calculation
of weak effects originating from virtual corrections. In fact, one should 
recall that real weak bosons are unstable and decay into high
transverse momentum leptons and/or jets, which are normally
captured by the detectors. In the definition of an
exclusive cross section then,
one tends to remove events with such additional particles.
Under such circumstances,
the (virtual) exchange of $Z$-bosons also generates similar logarithmic
corrections, 
$\alpha_{\mathrm{W}}\log^2(\sqrt{\hat{s}}/M_{Z}^2)$.
Besides, the genuinely weak contributions can  be
isolated in a gauge-invariant manner from purely EM effects,
at least in some simpler cases -- which do include the processes discussed
hereafter -- and the latter may or may not
be included in the calculation, depending on the observable being studied. 

A further aspect that should be recalled is that weak corrections naturally
introduce parity-violating effects in observables, detectable through
asymmetries in the cross-section, which are often regarded as an indication
of physics beyond the Standard Model
(SM) \cite{reviews,Maina:2003is,Dittmar:2003ir}. 
These effects are further enhanced if polarisation
of the incoming beams is exploited, such as at RHIC-Spin
\cite{Bourrely:1990pz,Ellis:2001ba} or a future Linear Collider(LC).
Comparison of theoretical predictions 
involving parity-violation with experimental data 
is thus used as another powerful tool for confirming or 
disproving  the existence of some beyond the SM scenarios, such as those 
involving right-handed weak currents \cite{Taxil:1997kj}, contact interactions
\cite{Taxil:1996vf} and/or new massive gauge bosons 
\cite{Taxil:1998ni,Taxil:1996vs}.

In view of all this,  it becomes of crucial importance to assess
the quantitative relevance of weak corrections.

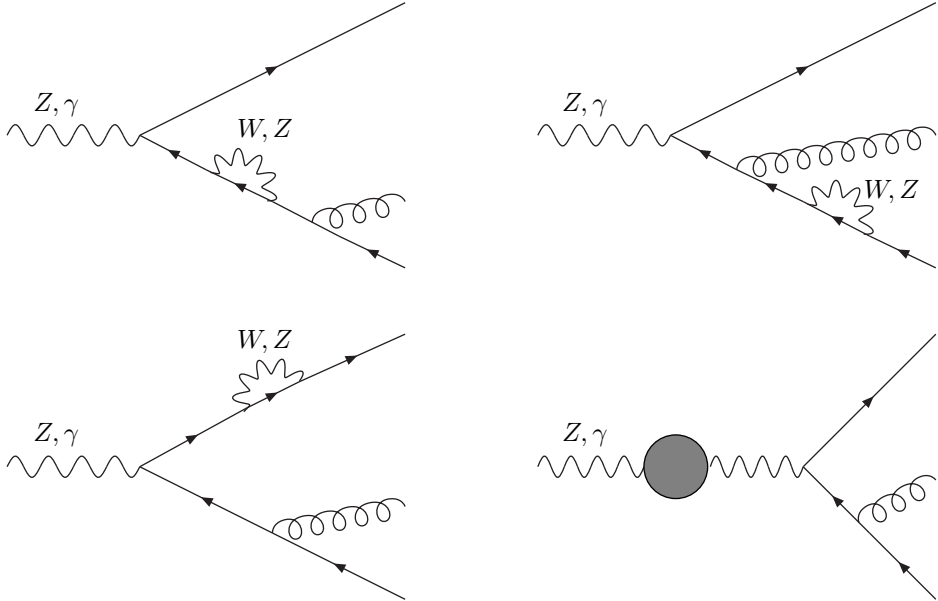
\begin{figure}[!bh]
\vspace*{1.cm}
\begin{center}
\begin{picture}(365,225)
\Photon(0,175)(50,175){4}{4}
\ArrowLine(150,125)(125,137) \Line(125,137)(100,150)
\ArrowLine(100,150)(75,162) \ArrowLine(75,162)(50,175)
\ArrowLine(50,175)(150,225) \Gluon(115,142)(150,150){4}{3}
\PhotonArc(87,155)(12,337,150){3}{5} \put(87,175){$W,Z$}
\put(10,185){$Z,\gamma$}

\Photon(200,175)(250,175){4}{4}
\ArrowLine(350,125)(325,137) \ArrowLine(325,137)(300,150)
\ArrowLine(300,150)(275,162) \ArrowLine(275,162)(250,175)
\ArrowLine(250,175)(350,225) \Gluon(275,162)(350,175){4}{8}
\PhotonArc(312,143)(12,334,150){3}{5} \put(324,152){$W,Z$}
\put(210,185){$Z,\gamma$}

\Photon(0,50)(50,50){4}{4}
\ArrowLine(150,0)(100,25) \ArrowLine(100,25)(50,50)
\ArrowLine(50,50)(90,72) \ArrowLine(90,72)(110,82)
\ArrowLine(110,82)(150,100) \Gluon(100,25)(150,35){4}{5}
\PhotonArc(100,76)(12,32,206){3}{5} \put(87,95){$W,Z$}
\put(10,60){$Z,\gamma$}

\Photon(200,50)(240,50){4}{4} \Photon(265,50)(300,50){4}{4}
\ArrowLine(350,0)(325,25) \ArrowLine(325,25)(300,50)
\ArrowLine(300,50)(350,100) \Gluon(321,29)(350,45){4}{3}
\GCirc(252,50){12}{.5}
\put(210,60){$Z,\gamma$}

\end{picture}
\vspace{1.0truecm}
\caption{Self-energy insertion graphs. The shaded blob on the incoming
wavy line represents all the contributions to the gauge boson 
self-energy and is dependent on the Higgs mass (hereafter, we will use
$M_H=115$ GeV for the latter). In this and all subsequent
figures the graphs in which the exchanged gauge boson is a $W$-boson
is accompanied by corresponding graphs in which the  $W$-boson is replaced
by its corresponding Goldstone boson. Since the Yukawa couplings are
proportional to the fermion masses, such graphs are only significant
 in the case of $b$-quark jets. There is a similar set of diagrams
in which the direction of the fermion line is reversed.}
\label{se_graphs}
\end{center}
\end{figure}

\section{Calculation}

Since we are considering weak corrections that can be
identified via their induced parity-violating effects and since we wish to
apply our results to the case of polarised electron and/or positron  
beams, it is convenient to work in terms of helicity matrix elements
(MEs). Thus, we define the helicity amplitudes 
${\cal A}^{(G)}_{\lambda_1, \lambda_2, \sigma}$ 
for the process
\be
V(q_1) + q(p_1) \rightarrow g(q_2) + q(p_2),
\ee
the scattering of a gauge boson of type $G$ (hereafter,
a possibly virtual photon $\gamma^*$ or a $Z$-boson) of helicity $\lambda_1$  and a quark
with helicity $\sigma$
into a gluon with helicity $\lambda_2$ and a massless quark
with the same helicity $\sigma$.\footnote{Note 
that all interactions considered here preserve the helicity along
the fermion line, including
those in which Goldstone bosons appear inside the loop, since these either
 occur in pairs or involve a mass insertion on the fermion line.}. 
The EW boson can have a longitudinal polarisation
component, so that the helicity $\lambda_1$ can take three values,
 $\pm1, \, 0$, for both the $\gamma^*$ and $Z$ 
gauge vectors\footnote{These helicities,
wherein $\pm1(0)$ are(is) transverse(longitudinal), are defined in a 
frame in which the particle
is {\it not} at rest, so that a fourth possible polarisation in the direction
of its four-momentum is irrelevant since its contribution
vanishes by virtue of current conservation.},
whereas $\lambda_2$ and $ \sigma$ can only be equal to $\pm 1$.

The general form of these amplitudes may be written as
\be 
{\cal A}^{(G)}_{\lambda_1, \lambda_2, \sigma}
  \ = \ \bar{u}(p_2) \Gamma \frac{\left(1+\sigma \gamma^5\right)}{2} u(p_1)
\ee
where $p_1$ and $p_2$ are the momenta of the incoming and outgoing quark 
respectively and $\Gamma$ stands for a sum of strings
 of Dirac $\gamma-$matrices
with coefficients, which, beyond tree level, 
involve integrals over loop momenta.
Since the helicity $\sigma$ of the fermions is conserved the strings 
must contain an odd number of $\gamma-$matrices. Repeated use of the
% Chrystoffel
Chisholm 
identity\footnote{This identity is only valid in four dimensions.
In our case, where we do not have infrared (i.e., 
soft and collinear) divergences, 
it is a simple matter to isolate the ultraviolet divergent contributions,
which are proportional to the tree-level MEs,
 and handle them separately. However, in $d$ dimensions
one needs to account for the fact that there are $2^{d/4}$ 
helicity states for the fermions and $(d-2)$ for the gauge bosons.
The method described here will {not} correctly trap terms
proportional to $(d-4)$ in coefficients of divergent integrals. 
}
 means that $\Gamma$ can always be expressed in the
form
\be \Gamma \ = \ C_1 \, \gamma \cdot p_1 \ + \ 
C_2 \, \gamma \cdot p_2 \ + \ C_3 \, \gamma \cdot q_2 \ + \ 
C_4 \, \sqrt{Q^2} \, \gamma \cdot n , \label{helicityme1} \ee
where $q_2$ is the momentum of the outgoing gluon, $Q^2=q_1^2$ is
the square momentum of the gauge boson,
and  $n$ is a unit vector normal to the quark and gluon momenta, more
precisely:
\begin{equation}
n_\mu =\frac{1}{\sqrt{2~p_1\cdot p_2~p_1\cdot q_2~p_2\cdot q_2}}
\varepsilon_{\mu\nu\rho\sigma}p_1^\nu p_2^\rho q_2^\sigma.
\end{equation}
The coefficient functions $C_i$ depend on the helicities  
$\lambda_1, \ \lambda_2, \ \sigma$ as well all independent kinematical
invariants and on all the couplings and masses of particles that enter into
the relevant perturbative contribution to the amplitude. 

For massless fermions the MEs of the first two terms
of eq.~(\ref{helicityme1})
vanish, and we are left with
\be \ {\cal A}^{(G)}_{\lambda_1, \lambda_2, \sigma} = 
C_3 \, \bar{u}(p_2) \gamma \cdot q_2 \frac{\left(1+\sigma \gamma^5\right)}{2}
 v(p_1) \, + \, C_4 \sqrt{Q^2}
\, \bar{u}(p_2) \gamma \cdot n \frac{\left(1+\sigma \gamma^5\right)}{2}
 v(p_1), \nonumber 
\ee

The relevant coefficient functions $C_3$ and $C_4$ 
are scalar quantities and can be projected on a graph-by-graph basis
using the projections
\be 
C_3 \ = \ {\rm Tr} \left( \Gamma \gamma \cdot v
\frac{\left(1+\sigma \gamma^5\right)}{2}
 \right)
\ee  
where $v$ is the vector
\be
v^\mu  \ =  \ = \frac{1}{-u} p_1^\mu \, + \, \frac{1}{s} p_2^\mu
   - \frac{-t}{-u \, s} q_2^\mu
\ee
with $s=(p_1+q_1)^2,\quad t=(p_2-p_1)^2,\quad u=(q_2-p_1)^2$
and
\be C_4 \ = \ -\frac{1}{2\sqrt{Q^2}}{\rm Tr} \left( \Gamma \gamma \cdot n
\frac{\left(1+\sigma \gamma^5\right)}{2}
 \right)  .\ee 
The basic matrix elements read 
\be
\bar{u}(p_2,\sigma)( \gamma \cdot q_2 ) u(p_1,\sigma) 
 \ = \ \sqrt{-u \, s} 
\quad\quad 
\bar{u}(p_2,\sigma) ( \gamma \cdot n ) u(p_1,\sigma) 
 \ = \ - i \, \sigma \, \sqrt{-t \, s} 
\ee

\begin{figure}[!h]
\vspace*{1.cm}
\begin{center}
\begin{picture}(365,225)

\Photon(100,175)(150,175){4}{4}
\ArrowLine(250,125)(225,137) \ArrowLine(225,137)(200,150)
\ArrowLine(200,150)(175,162) \ArrowLine(175,162)(150,175)
\ArrowLine(150,175)(250,225) \Gluon(200,150)(250,170){4}{5}
\PhotonArc(200,150)(24,154,330){3}{8} \put(187,115){$W,Z$}
\put(110,185){$Z,\gamma$}

\Photon(0,50)(50,50){4}{4}
\ArrowLine(150,0)(125,12) \ArrowLine(125,12)(87,30)
 \ArrowLine(87,30)(50,50)
\ArrowLine(50,50)(87,70) \ArrowLine(87,70)(150,100)
 \Gluon(115,17)(150,25){4}{3}
\Photon(97,25)(97,75){3}{5} \put(102,50){$W,Z$}
\put(10,60){$Z,\gamma$}

\Photon(200,50)(250,50){4}{4}
\ArrowLine(350,0)(325,12) \ArrowLine(325,12)(297,25)
 \Photon(297,25)(250,50){3}{5}
\Photon(250,50)(297,75){-3}{5} \ArrowLine(297,75)(350,100)
 \Gluon(315,17)(350,25){4}{3}
\ArrowLine(297,25)(297,75) \put(275,73){$W$} \put(275,19){$W$}
\put(210,60){$Z,\gamma$}

\end{picture}
\vspace{1.0truecm}
\caption{Vertex correction  graphs.  Again, same considerations 
as in the previous figure apply for the case of Goldstone bosons and
there is a similar set of graphs
in which the direction of the fermion line is reversed} \label{vertex_graphs}
\end{center}
\end{figure}
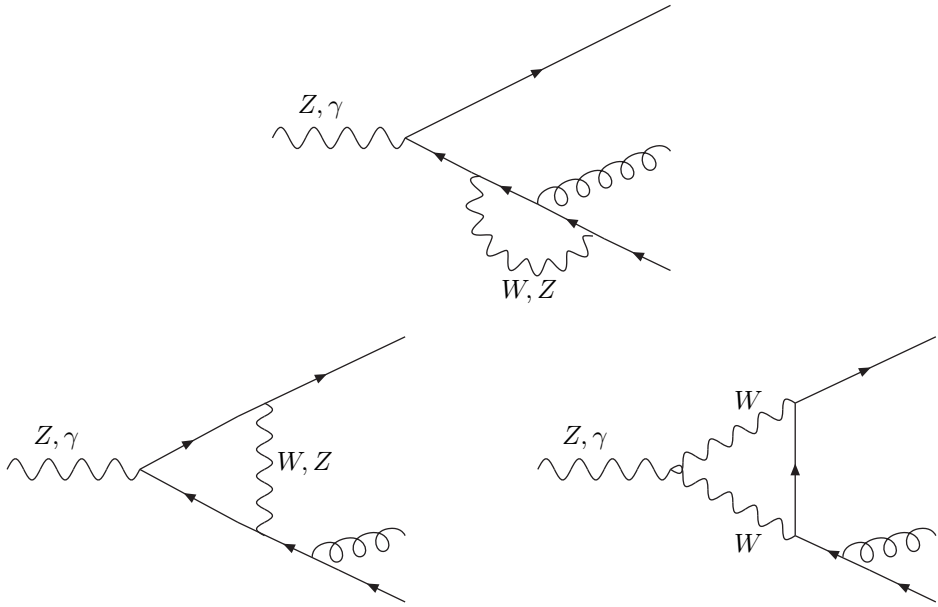

\begin{figure}[!bh]
\vspace*{1.cm}
\begin{center}
\begin{picture}(365,105)

\Photon(0,50)(50,50){4}{4}
\put(10,60){$Z,\gamma$}
\ArrowLine(150,29)(116,38) \ArrowLine(116,38)(83,44) \ArrowLine(83,44)(50,50)
\ArrowLine(50,50)(116,83) \ArrowLine(116,83)(150,100)
\Gluon(83,44)(150,0){-4}{8} \Photon(116,38)(116,83){4}{5}
\put(122,55){$W,Z$}

\Photon(200,50)(250,50){4}{4}
\ArrowLine(350,0)(307,20) 
 \Photon(307,20)(250,50){3}{5}
\Photon(250,50)(307,80){-3}{5} \ArrowLine(307,80)(350,100)
 \Gluon(307,50)(350,50){4}{5}
\ArrowLine(307,20)(307,50) \ArrowLine(307,50)(307,80)
 \put(275,73){$W$} \put(275,19){$W$}
\put(210,60){$Z,\gamma$}

\end{picture}
\vspace{1.0truecm}
\caption{Box  graphs. Again, same considerations as in the previous 
two figures apply for the case of Goldstone bosons. Here, the first 
graph is accompanied by a similar graph with the direction of the 
fermion line reversed whereas for the second
graph this reversal does {not} lead to a distinct
Feynman diagram.} \label{box_graphs}
\end{center}
\end{figure}
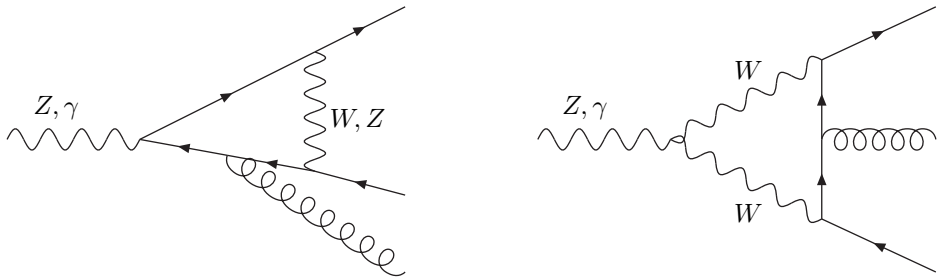

At one-loop level such helicity amplitudes   
acquire higher order corrections from the self-energy insertions on the
fermions and gauge bosons shown in Fig.~\ref{se_graphs},
from the vertex corrections shown in Fig.~\ref{vertex_graphs}
and from the box diagrams shown in Fig.~\ref{box_graphs}. As we have neglected 
here the masses of the external quarks, such higher order corrections
depend on the ratio $Q^2/M_{{W}}^2$, where $Q^2$ is the square
momentum of the gauge boson, as well as the EM coupling
constant $\alpha_{\rm EM}$ and the weak mixing angle $s_W\equiv\sin\theta_W$
(with $\alpha_{\rm EW}=\alpha_{\rm EM}/s_W^2$).
Furthermore, in the case where the external fermions are $b$-quarks, 
the loops involving the exchange of a $W$-boson lead to effects of
virtual $t$-quarks, so that the corrections also depend on 
the ratio $m_t^2/M_{{W}}^2$. (It is only in this case that the graphs 
involving
the exchange of the Goldstone bosons associated with the $W$-boson graphs
are relevant.)

The self-energy and vertex correction graphs contain ultraviolet divergences.
These have been subtracted using the `modified' Minimal Subtraction
($\MSbar$) scheme at
the scale $\mu=M_Z$. Thus the couplings are taken to be
those relevant for such a subtraction: e.g., the EM coupling,
$\alpha_{\mathrm{EM}}$, has been taken to be $1/128$ at the above subtraction
point. 
Two exceptions to this
renormalisation scheme have been the following:
\begin{enumerate}
\item the self-energy insertions
on external fermion lines, which have been subtracted on mass-shell,
so that the external fermion fields create or destroy particle states
with the correct normalisation;
\item the mass renormalization of the $Z$-boson propagator, which has also been 
carried out on mass-shell, so that the $Z$ mass does indeed refer to the
physical pole-mass.
\end{enumerate}

All these graphs are infrared and collinear convergent so that they
 may be expressed in terms of Passarino-Veltman \cite{VP} functions
which are then evaluated numerically. The expressions for
 each of these diagrams 
have been calculated using FORM \cite{FORM} and checked by an
independent program based on FeynCalc \cite{FeynCalc}. For the numerical
evaluation of the scalar integrals we have relied on FF \cite{FF1.9}. 
A further check on our results has been carried out
by setting the polarisation vector of the photon proportional to its momentum
and verifying that in that case the sum of all one-loop diagrams
vanishes, as required by gauge invariance.
%The full expressions for the contributions from these graphs are too
%lengthy to be reproduced here.

\section{Results}

\subsection{Factorisable Corrections to Three-Jet Production in 
Electron-Positron Annihilations\cite{EW3j}}

\begin{figure}[!bh]
\vspace*{1.cm}
\begin{center}
\epsfig{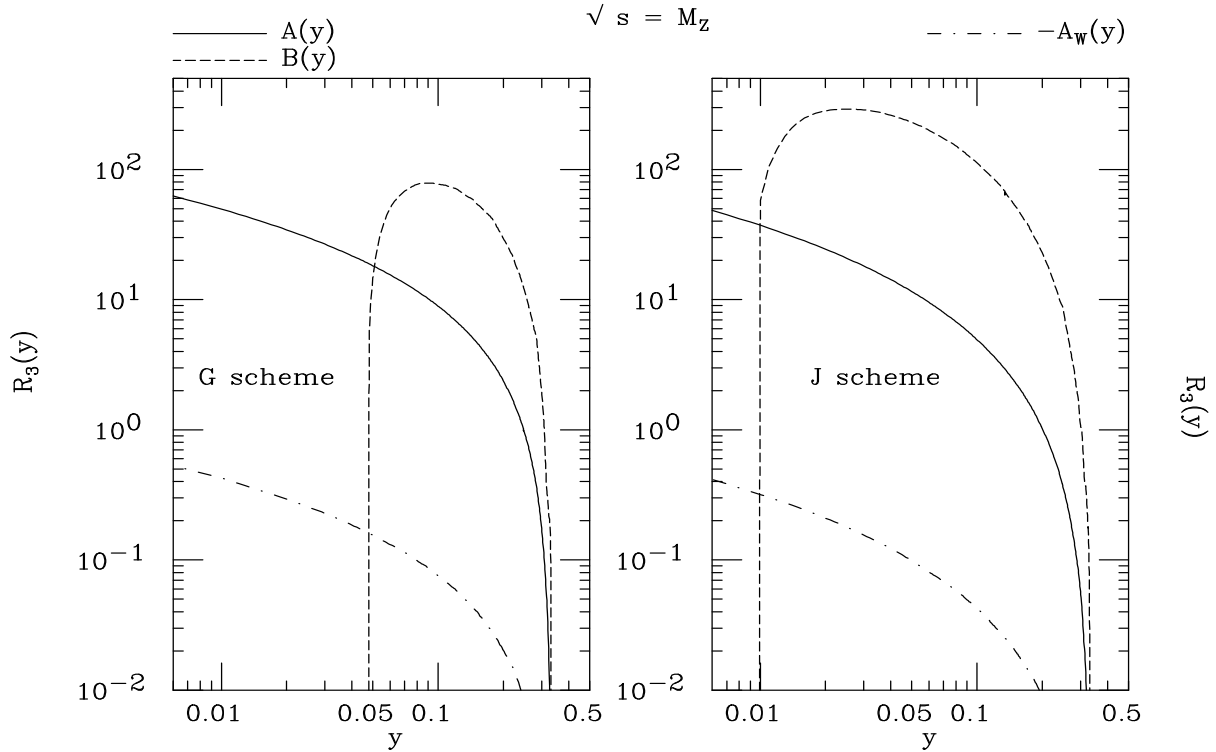}
\end{center}
\vskip -0.5cm
\caption{The $A(y)$, $-A_{\mathrm{W}}$ and $B(y)$ coefficient functions
of eqs.~(\ref{f3})--(\ref{f3EW}) for the Geneva and Jade
jet clustering algorithms, at $\sqrt s=M_Z$. (Notice that
the $\sim A_{\mathrm{W}}$ term
has been plotted with opposite sign
for better presentation.)}
\label{fig:y_LEP1}
\end{figure}

Here we report on the computation of 
one-loop weak effects entering three-jet production in electron-positron
annihilation
\begin{equation}\label{procj}
e^+e^-\to \gamma^*,Z\to \bar qqg\quad{\mathrm{(all~flavours)}},
\end{equation}
when no assumption is made on the flavour content of the final state,
so that a summation will be performed over $q=d,u,c,s,b$-quarks, and also
\begin{equation}\label{procb}
e^+e^-\to \gamma^*,Z\to \bar bbg,
\end{equation}
limited to the case of bottom quarks only in the final state. 
We restrict our attention to $\sqrt s=M_Z$
\footnote{See Ref.~\cite{2jet} for the corresponding weak corrections
to the Born process $e^+e^-\to\bar qq$ and Ref.~\cite{4jet} for the
$\sim n_{\rm f}$ component of those to $e^+e^-\to \bar qqgg$ (where 
$n_{\rm f}$ represents the number of light flavours).
For two-loop results on the former, see \cite{Beenakker:2000kb}.},
when the higher order effects arise only from initial or final state
interactions. 
These represent the so-called `factorisable' corrections, i.e.,
those involving loops 
not connecting the initial leptons to the final quarks,
which are the dominant ones at $\sqrt s=M_Z$ (where the width 
of the $Z$ resonance provides a natural cut-off for off-shellness
effects). The remainder, `non-factorisable' corrections,
while being negligible at $\sqrt s=M_{Z}$, 
are expected to play a quantitatively relevant role as $\sqrt s$ grows
larger. 
As a whole,
one-loop weak effects will become comparable to QCD ones
 at future LCs running at TeV energy scales\footnote{For example, 
at one-loop level,
in the case of the inclusive cross-section of $e^+e^-$
into hadrons, the QCD corrections are of  ${\cal O}
(\frac{\alpha_{\mathrm{S}}}{\pi})$, whereas
the EW ones are of ${\cal O}(\frac{\alpha_{\mathrm{EW}}}{4\pi}\log^2
\frac{s}{M^2_{{W}}})$, where $s$ is the collider CM energy
squared, so that at $\sqrt s=1.5$ TeV the former are identical to the latter,
of order 9\% or so.}. In contrast, 
at the $Z$ mass peak, where no logarithmic
enhancement occurs, one-loop weak effects are expected to appear
at the percent level, hence being of limited relevance at
LEP1 and SLC, where the final error on $\alpha_{\mathrm{S}}$
is of the same order or larger \cite{Dissertori}, but of crucial importance
at a GigaZ stage of a future LC, where the relative accuracy
of $\alpha_{\mathrm{S}}$ measurements is expected to be at the
$0.1\%$ level or smaller \cite{Winter}.

\begin{figure}[!bh]
\vspace*{1.cm}
\begin{center}
\hskip -0.75cm\epsfig{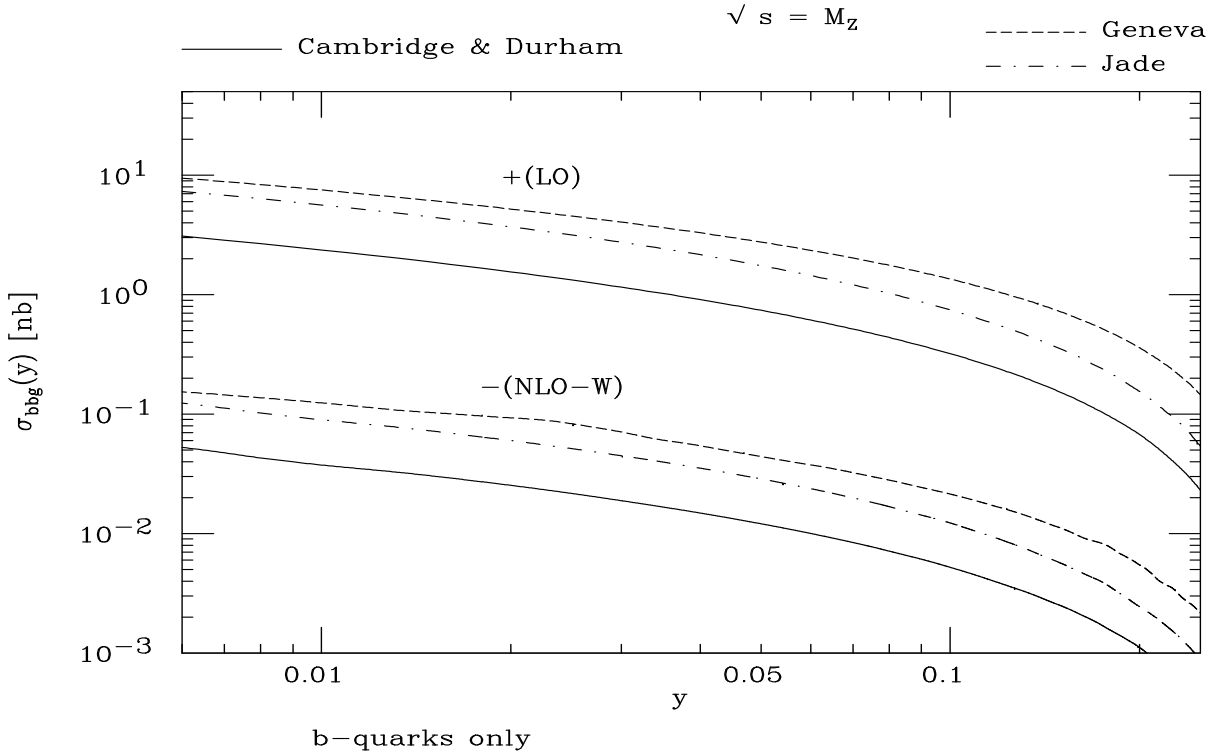}
\end{center}
\vskip -0.5cm
\caption{The total cross section for process (\ref{procb}) at LO
and NLO-W for the Cambridge, Durham, Geneva and
Jade jet clustering algorithms, at $\sqrt s=M_Z$. (Notice that the
NLO-W results have been plotted with opposite sign
for better presentation.)}
\label{fig:y_LEP1_b}
\end{figure}

For the choice $\mu=\sqrt s$ of the renormalisation scale, 
one can conveniently write the three-jet fraction in the following form:
\be
\label{f3}
R_3(y) =     \left( \frac{\as}{2\pi} \right)    A(y)
           + \left( \frac{\as}{2\pi} \right)^2  B(y) + ... ,
\ee
where the coupling constant $\as$ and the functions $A(y)$ and $B(y)$ 
are defined in  the $\overline{\mbox{MS}}$ scheme. An experimental fit
of the $R_n(y)$ jet fractions to the corresponding 
theoretical prediction is a powerful
way of determining $\as$ from multi-jet rates.
The weak corrections of interest (hereafter,
labelled as NLO-W) only contribute to three-parton final states. Hence,
in order to account for the latter, it will suffice to make the replacement
\be
\label{f3EW}
A(y)\to A(y)+A_{\mathrm{W}}(y)
\ee
in eq.~(\ref{f3}).

Fig.~\ref{fig:y_LEP1} displays the $A(y)$, $-A_{\rm{W}}(y)$ 
and $B(y)$ coefficients 
entering eqs.~(\ref{f3})--(\ref{f3EW}), as a function of 
$y(\equiv y_{\rm{cut}})$ for the Geneva(G) and Jade(J) 
jet algorithms at $\sqrt s=M_Z$.
%Notice that
%the sign of $A_{\mathrm{W}}$ has been reversed for better presentation.
A comparison
between $A(y)$ and $A_{\rm{W}}(y)$ reveals that the NLO-W corrections are
negative and remain
indeed at the percent level, i.e., of order $\frac{\aem}{2\pi s_W^2}$
without any logarithmic enhancement (since $\sqrt s\approx M_{{W}}, M_Z$).
They give rise to corrections to $\sigma_3(y)$ of --1\%, and
thus are generally much smaller than the NLO-QCD ones. In this context, no
systematic difference is seen with respect to the choice of jet clustering
algorithm, over the typical range of application of the latter at $\sqrt
s=M_Z$ (say $\ycut\gsim0.01$ for the G and J scheme).

As already mentioned, it should
now be recalled that jets originating from $b$-quarks can efficiently be
distinguished from light-quark jets. 
Besides, the $b$-quark component of the full three-jet sample is the
only one sensitive to $t$-quark loops in all diagrams of 
Figs.~\ref{se_graphs}--\ref{box_graphs}, hence one may expect somewhat
different effects from weak corrections to process (\ref{procb})
than to (\ref{procj}) (the residual dependence on the $Z \bar q q$
couplings is also different). This is confirmed by 
Fig.~\ref{fig:y_LEP1_b}, where we present the total cross section at $\sqrt
s=M_Z$ for $e^+e^-\to\gamma^*,Z\to\bar bbg$ as obtained at LO and NLO-W, for
our usual choice of jet clustering algorithms and separations. A close
inspection of the plots reveals that NLO-W effects can reach the 
$\sim -2.0\%$ level or so.

\begin{figure}[!bh]
\vspace*{1.cm}
\begin{center}
\epsfig{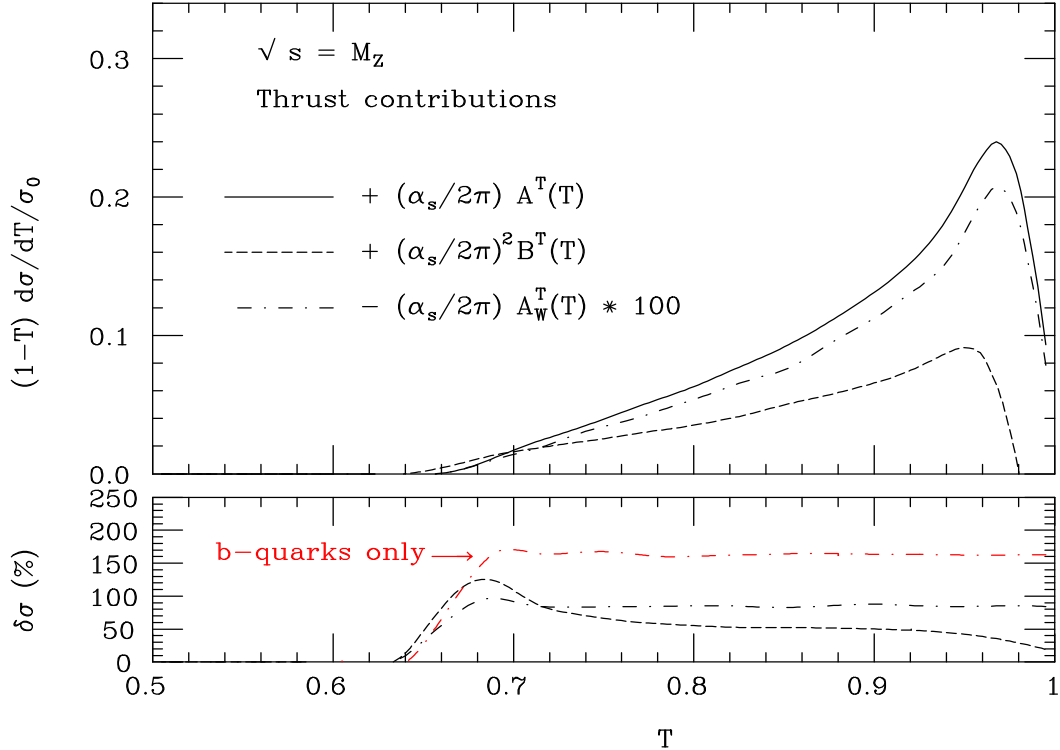}
\end{center}
\vskip -0.5cm
\caption{The LO, NLO-QCD and NLO-W  
contributions to the coefficient functions entering the
integrated Thrust distribution, see eq.~(\ref{T}), for
process (\ref{procj}) (top) and the relative 
size of the two NLO corrections (bottom), at $\sqrt s=M_Z$.
The correction for the case of $b$-quarks only is also
presented, relative to the LO results for process (\ref{procb}).
 (Notice that the
$\sim A_{\mathrm{W}}$ terms
have been plotted with opposite sign and
multiplied by hundred 
for better presentation.)}
\label{fig:thrust}
\end{figure}

In view of these percent effects being well above the error estimate
expected at a future high-luminosity LC running at the $Z$ pole,
it is then worthwhile to further consider the effects
of NLO-W corrections to some other `infrared-safe' jet observables typically
used in the determination
of $\as$, the so-called `shape variables'. 
A representative quantity in this respect is the Thrust (T)
distribution. This is defined as the sum of 
the longitudinal momenta relative to the (Thrust) axis $n_{\rm T}$ chosen
to maximise this sum, i.e.:
\begin{equation}\label{thrust}
\thrust = {\rm max} \frac{\sum_i |\vec{p_i}\cdot\vec{n_{\mathrm{T}}}|}
                         {\sum_i |\vec{p_i}|} ,
\end{equation} 
where $i$ runs over all final state clusters.
This quantity is identically one at Born level, getting
the first non-trivial
contribution through ${\cal O}(\as)$ from events of the
type (\ref{procj})--(\ref{procb}). Also notice that any other higher 
order contribution will affect this observable. Through ${\cal O}(\as^2)$,
for the choice $\mu =\sqrt s$ of the renormalisation scale, 
the T distribution can be parametrised in the following form:
\begin{equation}\label{T}
(1-{\rm{T}})\frac{d\sigma}{d\thrust}\frac{1}{\sigma_0} = 
\left(\frac{\as}{2\pi}\right)   A^{\Thrust}(\thrust)+
\left(\frac{\as}{2\pi}\right)^2 B^{\Thrust}(\thrust).
\end{equation} 
Again, the replacement 
\begin{equation}\label{TW}
A^{\Thrust}(\thrust)\to A^{\Thrust}(\thrust)+A^{\Thrust}_{\rm{W}}(\thrust) 
\end{equation} 
accounts for the inclusion of the NLO-W contributions.

We plot the terms $\left(\frac{\as}{2\pi}\right)A^{\Thrust}(\thrust)$,
$\left(\frac{\as}{2\pi}\right)A^{\Thrust}_{\rm{W}}(\thrust)$ and 
$\left(\frac{\as}{2\pi}\right)^2B^{\Thrust}(\thrust)$ in Fig.~\ref{fig:thrust},
always at  $\sqrt s=M_Z$,
alongside the relative rates of the NLO-QCD and NLO-W terms 
with respect to the LO contribution. Here, it can be seen that
the NLO-W effects can reach the level of $-1\%$ or so and that they are
fairly constant for $0.7\lsim {\rm{T}}\lsim 1$. For the case of $b$-quarks
only, similarly to what seen already for the inclusive
rates, the NLO-W corrections are larger, as they can reach the $-1.6\%$ level.

\subsection{$Z$/$\gamma$ Hadroproduction at Finite Transverse
Momentum\cite{EWZj}}

\begin{figure}[!bh]
\vspace{1.cm}
\begin{center}
{\epsfig{file=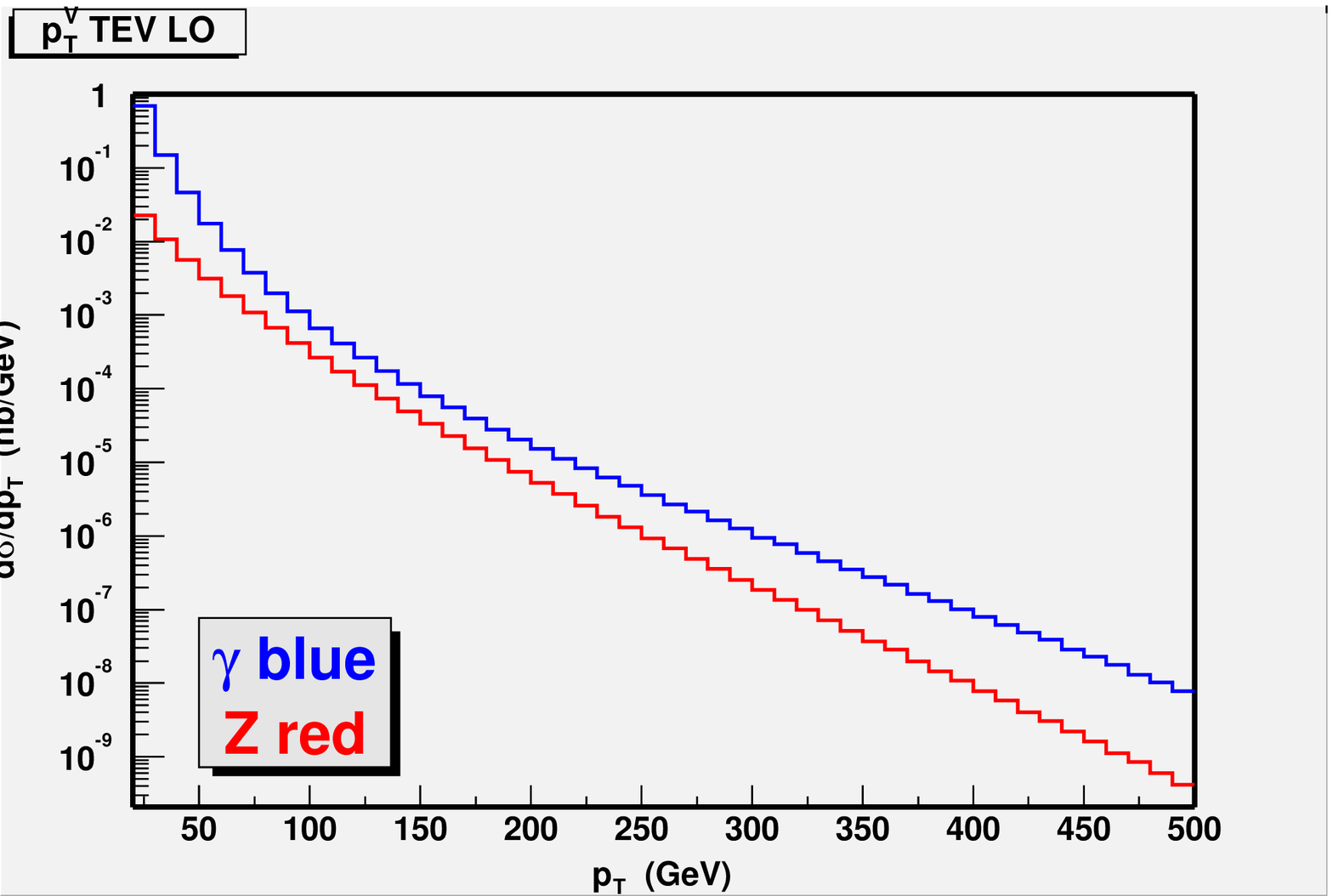,  width=10cm, angle=0}}
{\epsfig{file=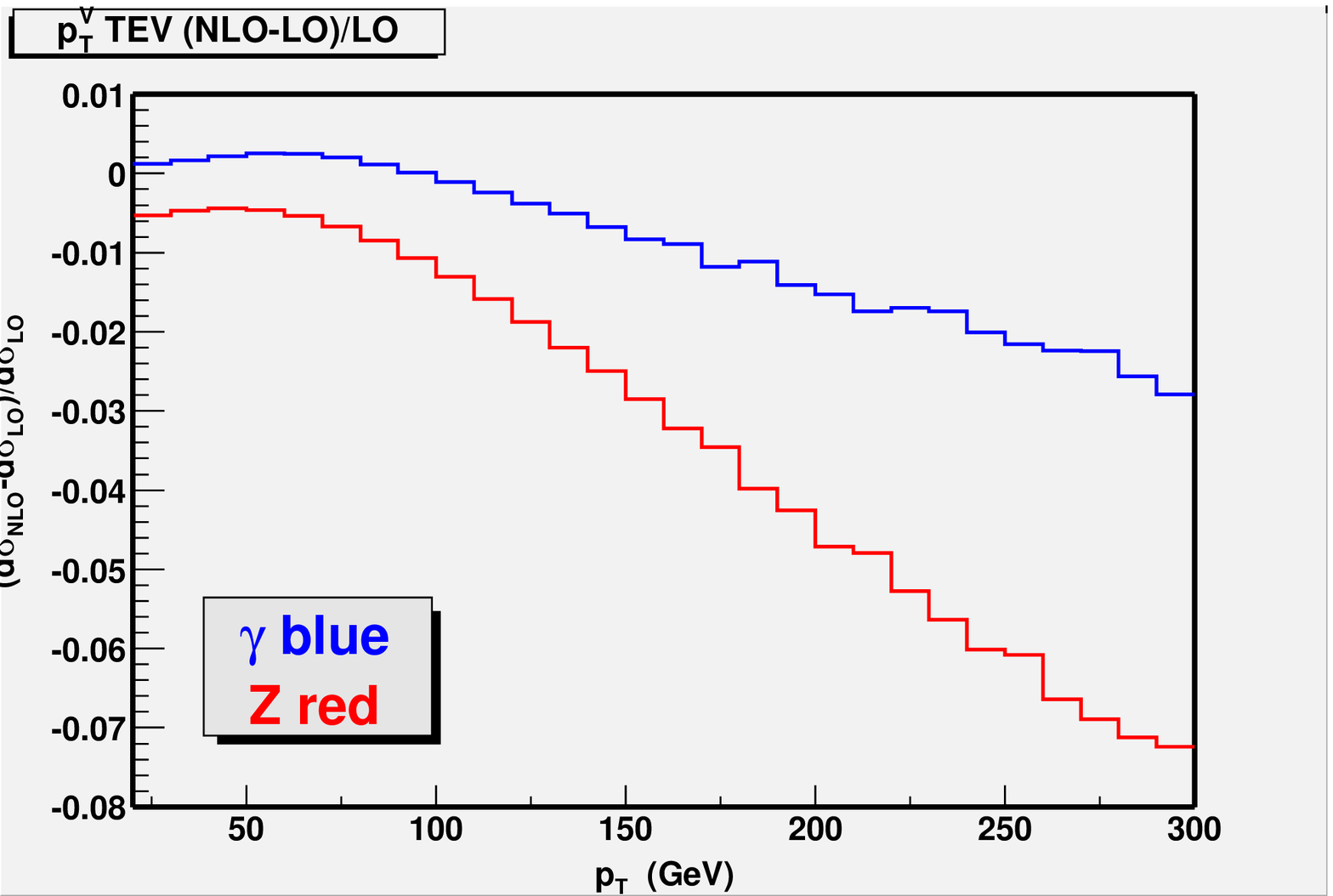, width=10cm, angle=0}}
\end{center}
\vspace*{-0.75cm}
\caption{\small The transverse momentum dependence of the $\gamma$- and
$Z$-boson cross 
sections in (\ref{procs_neutral}) at LO 
(top frame) and the size of the one-loop weak 
corrections (bottom frame), at Tevatron ($\sqrt s_{p\bar p}=2$ TeV).
Notice that the pseudorapidity range of the jet
in the final state is limited to $|\eta|<3$.} 
\label{fig:pT-Tev}
\end{figure}

The neutral-current processes ($V=\gamma,Z$)
\begin{equation}\label{procs_neutral}
q\bar q \to g V\quad{\rm{and}}\quad q(\bar q) g\to q(\bar q) V
\end{equation}
with $V\to\ell^+\ell^-$
are two of the cleanest probes of the partonic content of (anti)protons,
in particular of antiquark and gluon
densities. In order to measure the latter it is necessary to study
the vector boson $p_T$ spectrum. According to \cite{bego1,bego2} the gluon
density dominates for $p_T\ > Q/2$ where $Q$ is the lepton pair 
invariant mass.
In the presence of polarised beams these reactions give access to the
spin--dependent gluon distribution which is presently only poorly known.
Thanks to the introduction of improved algorithms 
\cite{Frixione:1999pl}--\cite{Frixione:1999ya} for the 
selection of (prompt) photons generated in the hard scatterings  
(\ref{procs_neutral}), 
as opposed to those generated in the fragmentation of the accompanying
gluon/quark jet, and to the high experimental resolution achievable
in reconstructing 
$Z\to\ell^+\ell^-$ ($\ell=e,\mu$) decays, they are regarded -- together
with the twin charged-current channels 
\begin{equation}\label{procs_charged}
q\bar q' \to g W \quad{\rm{and}}\quad q(\bar q) g\to q'(\bar q')W,
\end{equation}
wherein $W\to\ell\nu_\ell$ -- as precision observables in hadronic 
physics. In fact, in some instances, accuracies
of order one percent are expected to be 
attained in measuring these processes \cite{reviews},
both at present and future proton-(anti)proton 
experiments. These include the Relativistic Heavy Ion Collider running
with polarised proton beams (RHIC-Spin) at BNL ($\sqrt s_{pp}=300-600$ GeV), 
the Tevatron collider at FNAL (Run 2, $\sqrt s_{p\bar p}=2$ TeV) 
and the Large Hadron Collider (LHC) at CERN
($\sqrt s_{pp}=14$ TeV).

Not surprisingly then, a lot of effort has been spent over the years
in computing higher order corrections to all such Drell--Yan type processes.
To stay with the
neutral-current ones these include 
next-to-leading order (NLO) QCD
calculations of both prompt-photon \cite{coka,govo} and vector boson
production \cite{kamal}. QCD corrections to the $p_T$ distributions have been
computed in Refs.~\cite{elma,arre}.
As for the full $\cal O (\alpha)$ Electro-Weak (EW)
corrections to $Z$ production and continuum
neutral-current processes (at zero transverse momentum), these have been 
completed in \cite{Baur:2001ze} (see also \cite{Haywood:1999qg}), building on
the calculation of the QED part in \cite{Baur:1998zf}.

\begin{figure}[!bh]
\vspace*{1.cm}
\begin{center}
{\epsfig{file=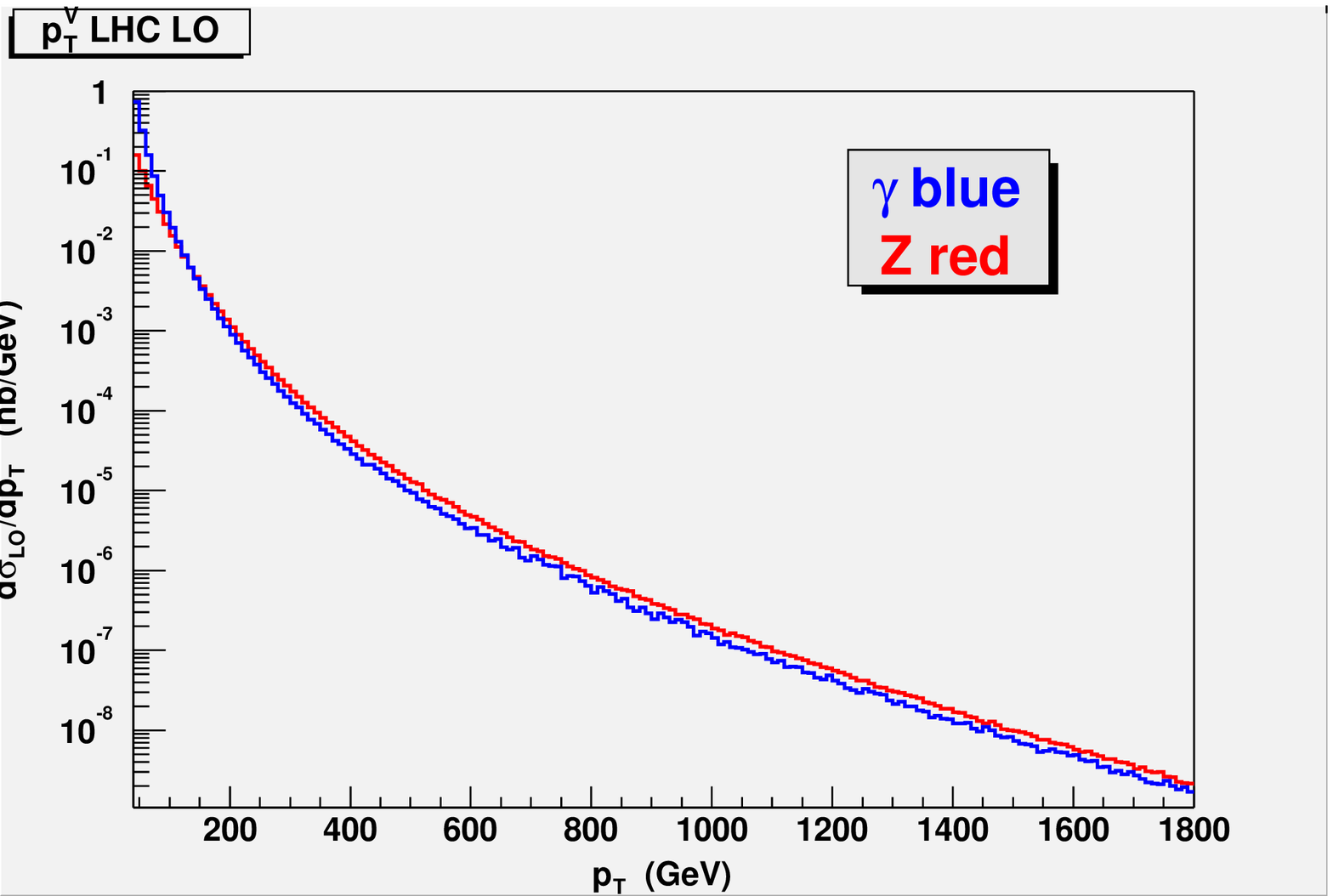,  width=10cm, angle=0}}
{\epsfig{file=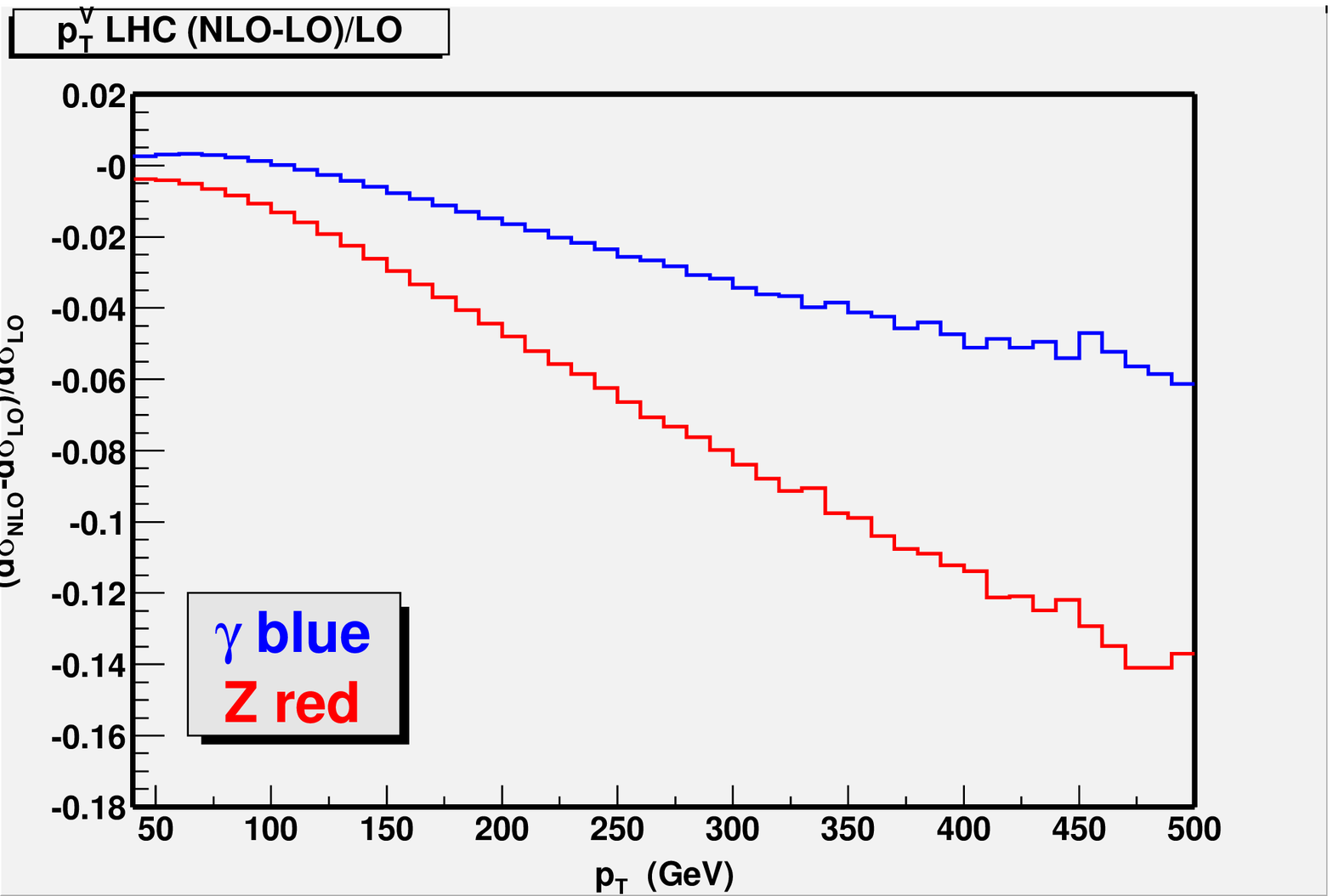, width=10cm, angle=0}}
\end{center}
\vspace*{-0.75cm}
\caption{\small The transverse momentum dependence of the $\gamma$- and
$Z$-boson cross 
sections in (\ref{procs_neutral}) at LO 
(top frame) and the size of the one-loop weak 
corrections (bottom frame), at LHC ($\sqrt s_{pp}=14$ TeV).
Notice that the pseudorapidity range of the jet
in the final state is limited to $|\eta|<4.5$.} 
\label{fig:pT-LHC}
\end{figure}

Figs.~\ref{fig:pT-Tev}--\ref{fig:pT-LHC} show the effects of the 
${\cal O}(\alpha_{\rm{S}}\alpha_{\rm{EW}}^2)$
terms relatively to the ${\cal O}(\alpha_{\rm{S}}\alpha_{\rm{EW}})$
Born results ($\alpha_{\rm{EM}}$ replaces $\alpha_{\rm{EW}}$ for photons),
as well as the absolute magnitude of the latter, as a function
of the transverse momentum, at Tevatron and LHC, respectively. 
The corrections are found to be rather large at both colliders, particularly
for $Z$-production. In case of the latter,
such effects are of order --7\% at Tevatron for $p_T\approx 300$ GeV
and --14\% at LHC for $p_T\approx 500$ GeV. In general, above 
$p_T\approx100$ GeV,
they tend to (negatively) increase, more or less linearly, with $p_T$.
Such effects will be hard to observe at Tevatron but
will indeed be observable at LHC. 
For example, at FNAL, for $Z$-production and decay into electrons and muons
with BR$(Z\rightarrow e,\mu)\approx 6.5\%$, assuming
$L= 2-20$ fb$^{-1}$ as integrated luminosity, in
a window of 10 GeV at $p_T = 100$ GeV, one finds
500--5000 $Z+j$ events at LO, hence a
$\delta\sigma/\sigma\approx -1.2\%$ EW NLO correction corresponds to only
6--60 fewer 
events. At CERN, for the same production and decay channel, assuming now 
$L= 30$ fb$^{-1}$, in a window of 40 GeV at $p_T = 450$ GeV, 
we expect about 2000 $Z+j$ events from LO, so that a 
$\delta\sigma/\sigma\approx -12\%$ EW NLO correction 
corresponds to 240 fewer events. In line with the normalisations seen
in the top frames of  Figs.~\ref{fig:pT-Tev}--\ref{fig:pT-LHC}
and the size of the corrections
in the bottom ones, absolute rates for the photon are similar
to those for the massive gauge boson while ${\cal O}(\alpha_{\rm{S}}\alpha_{\rm{EW}}^2)$ corrections are about a factor of two 
smaller.

\subsection{$b\overline{b}$ Production at TeV Energy Hadron Colliders\cite{Maina:2003is}}

\begin{figure}[!bh]
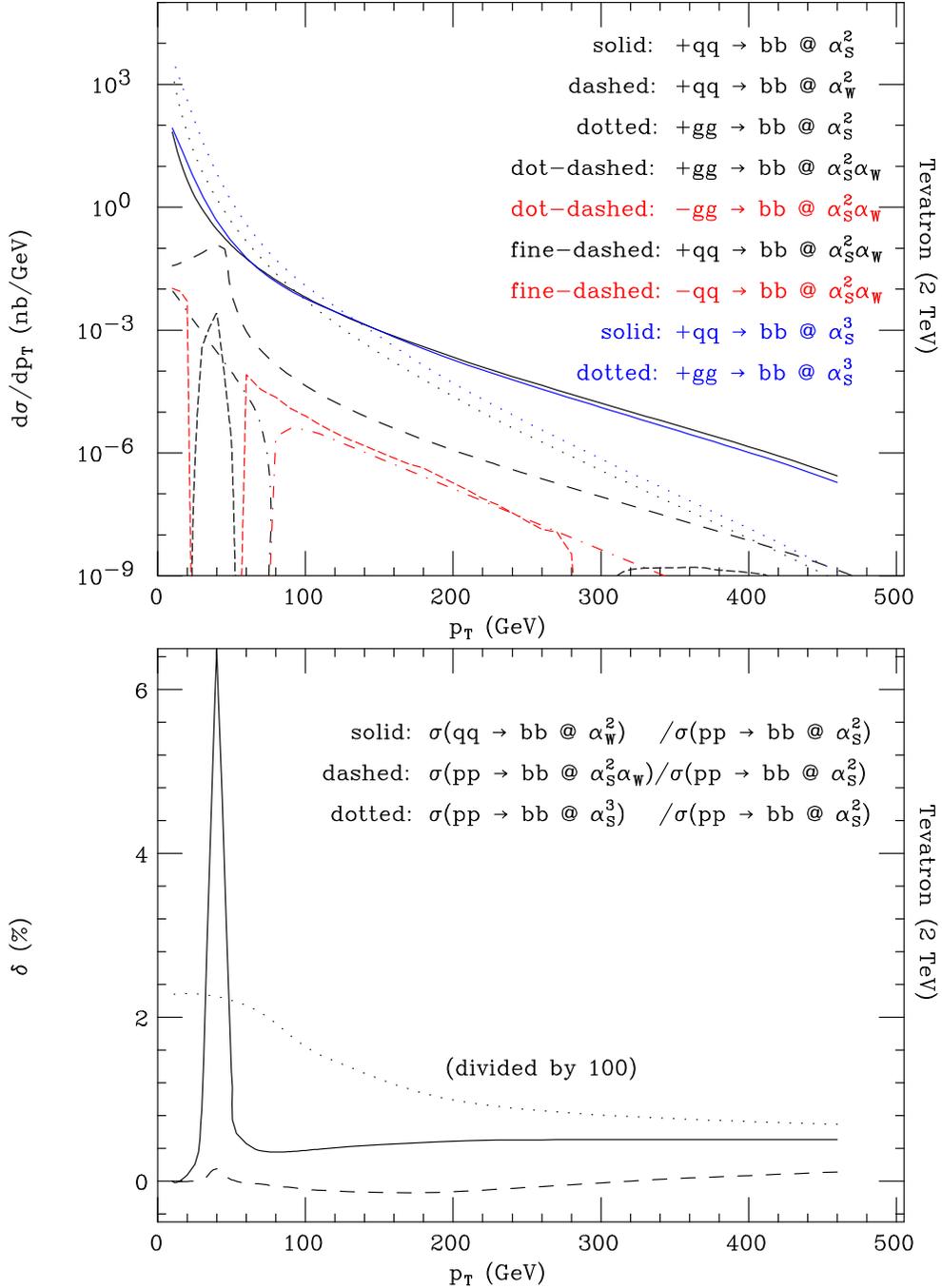

\begin{center}
\vspace{1.cm}
{\epsfig{file=sigmabb_Tev.ps, width=9cm, angle=90}}
%\vspace*{0.75cm}
{\epsfig{file=ratiobb_Tev.ps, width=9cm, angle=90}}
\end{center}
\vspace*{-0.75cm}
\caption{The total cross section contributions vs. the transverse momentum
of the $b$-jet for $p\bar p\to b\bar b$ production at
Tevatron (2 TeV) as obtained via the various subprocesses discussed in the
text (top) and the corrections due to the 
$\alpha_{\rm{EW}}^2$, $\alpha_{\rm{S}}^2\alpha_{\rm{EW}}^2$ 
and $\alpha_{\rm{S}}^3$ terms
relative to the $\alpha_{\rm{S}}^2$ ones (bottom).}
\label{sigmabb_Tev}
\end{figure}

\begin{figure}[!bh]
\begin{center}
\vspace{1.cm}
{\epsfig{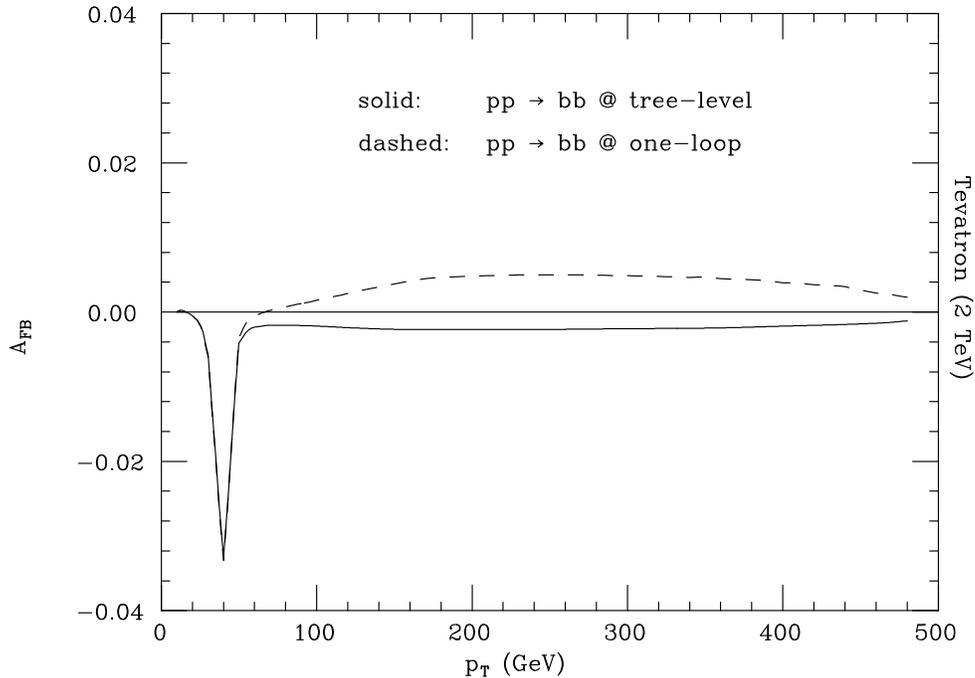}}
%\vspace*{0.75cm}
%{\epsfig{file=ratiobb_AFB_Tev.ps, width=9cm, angle=90}}
\end{center}
\vspace*{-0.75cm}
\caption{The forward-backward  
asymmetry vs. the transverse momentum
of the $b$-jet for $p\bar p\to b\bar b$ events at
Tevatron (2 TeV), as obtained at tree-level and one-loop order (top)
and the relative correction of the latter to the former (bottom).
(Errors in the ratio are statistical.)}
\label{sigmabb_AFB_Tev}
\end{figure}

\begin{figure}[!bh]
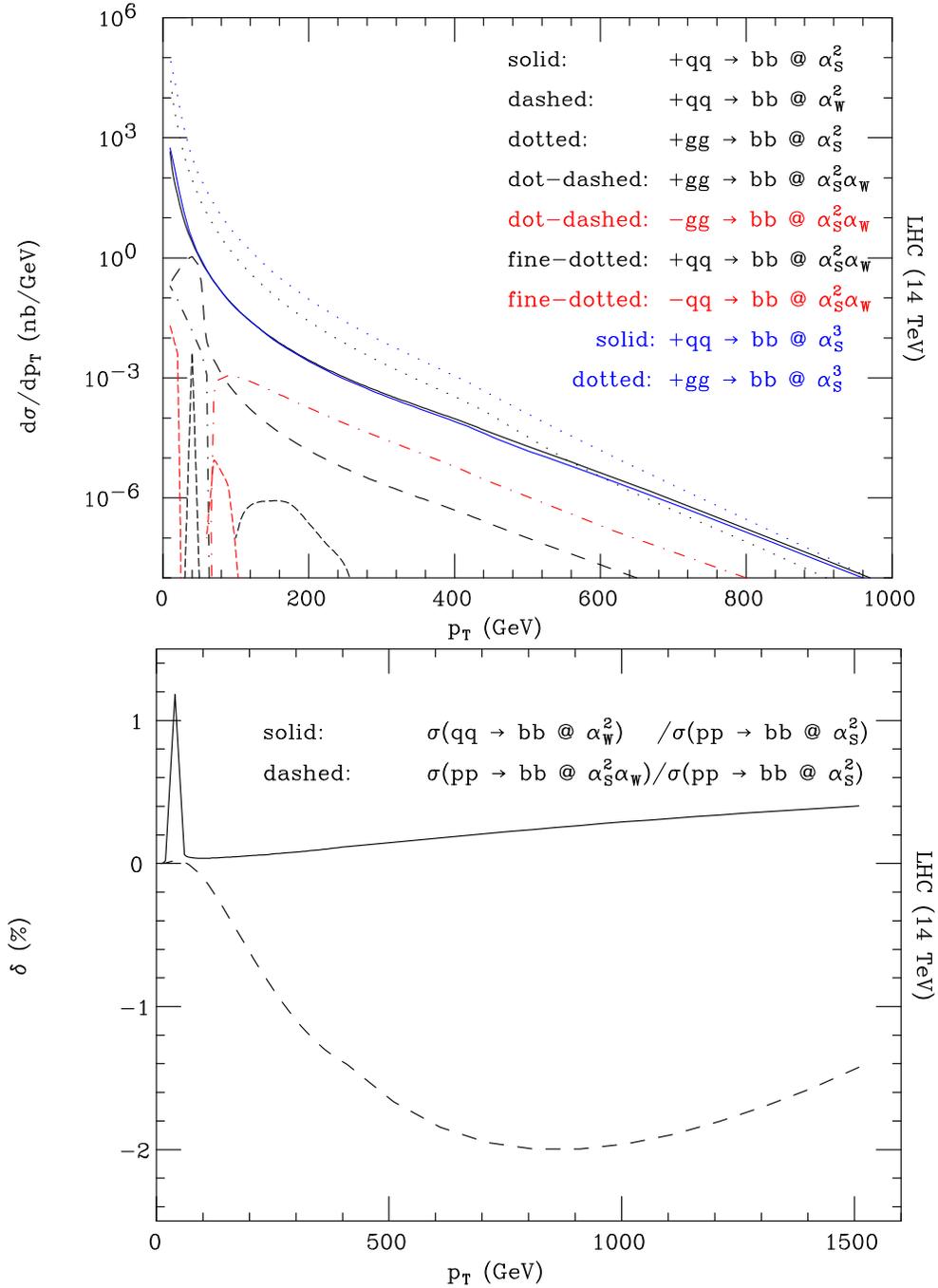

\begin{center}
\vspace{1.cm}
{\epsfig{file=sigmabb_LHC.ps, width=9cm, angle=90}}
%\vspace*{0.75cm}
{\epsfig{file=ratiobb_LHC.ps, width=9cm, angle=90}}
\end{center}
\vspace*{-0.75cm}
\caption{The total cross section contributions vs. the transverse momentum
of the $b$-jet for $pp\to b\bar b$ production at 
LHC (14 TeV) as obtained via the various subprocesses discussed in the
text (top) and the corrections due to the 
$\alpha_{\rm{EW}}^2$ and $\alpha_{\rm{S}}^2\alpha_{\rm{EW}}^2$ terms
relative to the $\alpha_{\rm{S}}^2$ ones (bottom). (Here,
we do not show the corrections due to $\alpha_{\rm{S}}^3$ terms as results are
perturbatively unreliable, given that $K$-factors as large as 3--4 can appear.)}
\label{sigmabb_LHC}
\end{figure}

The inclusive $b$-jet cross section at both Tevatron and LHC is dominated
by the pure QCD contributions
$gg\to b\bar b$ and $q\bar q\to b\bar b$, 
known through order $\alpha_{\rm S}^n$ for $n=2,3$.
%%%%%%%%%%%%%%%%%
%However, as already intimated previously, the pure QCD subprocesses
%contribute negligibly to spin asymmetries defined via a $b\bar b$
%final state, such as the forward-backward
%one recalled above. 
%%%%%%%%%%%%%%%%5
Of particular relevance in this context is the fact
that for the flavour creation mechanisms no 
$\alpha_{\rm S}\alpha_{\rm{W}}$ tree-level contributions are allowed,
because of colour conservation: i.e., 
\vspace*{0.95truecm}
\begin{eqnarray}
% \nonumber
\begin{picture}(300,-300)
\SetScale{1.0}
\SetWidth{1.2}
\SetOffset(0,-60)
\ArrowLine(30,90)(45,75)
\ArrowLine(45,75)(30,60)
\Gluon(45,75)(60,75){3}{3}
\Text(25,95)[]{\small $q$}
\Text(25,55)[]{\small $\bar q$}
\ArrowLine(60,75)(75,90)
\ArrowLine(75,60)(60,75)
\Text(80,95)[]{\small $b$}
\Text(80,55)[]{\small $\bar b$}
\Text(105,75)[]{$*~[$}
\ArrowLine(130,90)(145,75)
\ArrowLine(145,75)(130,60)
\Photon(145,75)(160,75){3}{3}
\Text(125,95)[]{\small $q$}
\Text(125,55)[]{\small $\bar q$}
\ArrowLine(160,75)(175,90)
\ArrowLine(175,60)(160,75)
\Text(180,95)[]{\small $b$}
\Text(180,55)[]{\small $\bar b$}
\Text(215.,75)[]{$]^\dagger~~=~~0$,}
\end{picture} 
\end{eqnarray}
%\vspace*{0.25cm}
\noindent
where the wavy line represents a $Z$ boson (or a photon) and the 
helical one a gluon. Tree-level asymmetric terms through 
the order $\alpha_{\rm{EW}}^2$ are however finite, as they
are given by non-zero quark-antiquark initiated
diagrams such as the one above wherein the gluon is
replaced by a $Z$ boson (or a photon).  The latter are the
leading contribution to the forward-backward asymmetry (more precisely,
those graphs containing one or two $Z$ bosons are, as those involving
two photons are subleading in this case, even with respect to
the pure QCD contributions).

We have computed one-loop and (gluon) radiative contributions through
the order $\alpha_{\rm S}^2\alpha_{\rm{W}}$, which -- in the case
of quark-antiquark induced subprocesses -- are represented 
schematically by the 
following diagrams:
%\newpage
\vspace*{0.95truecm}
\begin{eqnarray}
\label{as2aW-qq} \nonumber
\begin{picture}(300,-300)
\SetScale{1.0}
\SetWidth{1.2}
\SetOffset(0,-60)
\ArrowLine(30,90)(45,75)
\ArrowLine(45,75)(45,60)
\ArrowLine(45,60)(30,45)
\Gluon(45,75)(60,75){3}{3}
\Text(25,95)[]{\small $q$}
\Text(25,40)[]{\small $\bar q$}
\ArrowLine(60,75)(75,90)
\ArrowLine(60,60)(60,75)
\ArrowLine(75,45)(60,60)
\Gluon(45,60)(60,60){3}{3}
\Text(80,95)[]{\small $b$}
\Text(80,40)[]{\small $\bar b$}
\Text(105,75)[]{$*~[$}
\ArrowLine(130,90)(145,75)
\ArrowLine(145,75)(130,60)
\Photon(145,75)(160,75){3}{3}
\Text(125,95)[]{\small $q$}
\Text(125,55)[]{\small $\bar q$}
\ArrowLine(160,75)(175,90)
\ArrowLine(175,60)(160,75)
\Text(180,95)[]{\small $b$}
\Text(180,55)[]{\small $\bar b$}
\Text(245,75)[]{$]^\dagger~~ \ +  $ crossed box $+$}
\end{picture} 
\end{eqnarray}
\vspace*{0.75truecm}
\begin{eqnarray}
 \nonumber
\begin{picture}(300,-300)
\SetScale{1.0}
\SetWidth{1.2}
\SetOffset(0,-60)
\ArrowLine(30,90)(45,75)
\ArrowLine(45,75)(45,60)
\ArrowLine(45,60)(30,45)
\Photon(45,75)(60,75){3}{3}
\Text(25,95)[]{\small $q$}
\Text(25,40)[]{\small $\bar q$}
\ArrowLine(60,75)(75,90)
\ArrowLine(60,60)(60,75)
\ArrowLine(75,45)(60,60)
\Gluon(45,60)(60,60){3}{3}
\Text(80,95)[]{\small $b$}
\Text(80,40)[]{\small $\bar b$}
\Text(105,75)[]{$*~[$}
\ArrowLine(130,90)(145,75)
\ArrowLine(145,75)(130,60)
\Gluon(145,75)(160,75){3}{3}
\Text(125,95)[]{\small $q$}
\Text(125,55)[]{\small $\bar q$}
\ArrowLine(160,75)(175,90)
\ArrowLine(175,60)(160,75)
\Text(180,95)[]{\small $b$}
\Text(180,55)[]{\small $\bar b$}
\Text(245,75)[]{$]^\dagger~~\ +$ crossed box $+$}
\end{picture} 
\end{eqnarray}
\vspace*{0.75truecm}
\begin{eqnarray}
 \nonumber
\begin{picture}(300,-300)
\SetScale{1.0}
\SetWidth{1.2}
\SetOffset(0,-60)
\ArrowLine(30,90)(45,75)
\Photon(35,85)(35,65){3}{3}
\ArrowLine(45,75)(30,60)
\Gluon(45,75)(60,75){3}{3}
\Text(25,95)[]{\small $q$}
\Text(25,55)[]{\small $\bar q$}
\ArrowLine(60,75)(75,90)
\ArrowLine(75,60)(60,75)
\Text(80,95)[]{\small $b$}
\Text(80,55)[]{\small $\bar b$}
\Text(105,75)[]{$*~[$}
\ArrowLine(130,90)(145,75)
\ArrowLine(145,75)(130,60)
\Gluon(145,75)(160,75){3}{3}
\Text(125,95)[]{\small $q$}
\Text(125,55)[]{\small $\bar q$}
\ArrowLine(160,75)(175,90)
\ArrowLine(175,60)(160,75)
\Text(180,95)[]{\small $b$}
\Text(180,55)[]{\small $\bar b$}
\Text(265,75)[]{$]^\dagger~~+$ {\normalsize other three vertices~~+}}
\end{picture} 
\end{eqnarray}
\vskip0.5cm
\hskip2.70cm{\normalsize{+~all self-energies~+}}
\vskip-0.05cm
%\vskip1.0cm
\vspace*{0.75truecm}
\begin{eqnarray}
 %\nonumber
\begin{picture}(300,-300)
\SetScale{1.0}
\SetWidth{1.2}
\SetOffset(0,-60)
\ArrowLine(30,90)(45,75)
\Gluon(35,85)(45,95){3}{3}
\ArrowLine(45,75)(30,60)
\Gluon(45,75)(60,75){3}{3}
\Text(25,95)[]{\small $q$}
\Text(25,55)[]{\small $\bar q$}
\ArrowLine(60,75)(75,90)
\ArrowLine(75,60)(60,75)
\Text(80,95)[]{\small $b$}
\Text(80,55)[]{\small $\bar b$}
\Text(105,75)[]{$*~[$}
\ArrowLine(130,90)(145,75)
\Gluon(170,85)(155,95){3}{3}
\ArrowLine(145,75)(130,60)
\Photon(145,75)(160,75){3}{3}
\Text(125,95)[]{\small $q$}
\Text(125,55)[]{\small $\bar q$}
\ArrowLine(160,75)(175,90)
\ArrowLine(175,60)(160,75)
\Text(180,95)[]{\small $b$}
\Text(180,55)[]{\small $\bar b$}
\Text(260,75)[]{$]^\dagger~+$ {\normalsize{gluon permutations.}}~~}
\end{picture} 
\end{eqnarray}
The gluon bremsstrahlung graphs are needed in order
to cancel the infinities arising in the virtual contributions when
the intermediate gluon becomes infrared.
 Furthermore, one also has to include 
$\alpha_{\rm S}^2\alpha_{\rm{W}}$ terms induced by gluon-gluon
scattering, that is, interferences between the graphs displayed in
Fig.~1 of Ref.~\cite{Ellis:2001ba} and the tree-level ones for
$gg\to b\bar b$.

The total cross section,  $\sigma(p\bar p\to b\bar b)$,
for Tevatron (Run2) can be found in Fig.~\ref{sigmabb_Tev} (top), 
as a function of the transverse momentum of the $b$-jet (or
$\bar{b}$-jet) and decomposed in terms of 
the various subprocesses discussed so far.  (Hereafter, 
the pseudorapidity is limited between $-2$ and $2$ in the partonic CM frame.)
The dominance at inclusive
level of the pure QCD contributions is manifest, over the entire
$p_T$ spectrum. At low transverse momentum it is the gluon-gluon
induced subprocess that dominates, with the quark-antiquark one
becoming the strongest one at large $p_T$. The QCD $K$-factors, defined
as the ratio of the $\alpha_{\rm{S}}^3$ rates to the $\alpha_{\rm{S}}^2$ 
ones are rather large, of order 2 and positive for the $gg\to b\bar b$
subprocess and somewhat smaller  for
the $q\bar q\to b\bar b$ case, which has a $p_T$-dependent
sign\footnote{Further notice that
in QCD at NLO one also has (anti)quark-gluon induced (tree-level) 
contributions, which are of similar strength to those via 
gluon-gluon and quark-antiquark scattering but which have not been shown 
here.}. The tree-level $\alpha_{\rm{EW}}^2$ terms are much smaller
than the QCD rates, typically by three orders of magnitude, with the
exception of the $p_T\approx M_Z/2$ region, where one can appreciate
the onset of the $Z$ resonance in $s$-channel. All above terms are
positive. The $\alpha_{\rm{S}}^2\alpha_{\rm{EW}}$ subprocesses display
a more complicated structure, as their sign can change over the
transverse momentum spectrum considered, and the behaviour is different
in $q\bar q\to b\bar b(g)$ from $gg\to b\bar b$.
Overall, the rates for the $\alpha_{\rm{S}}^2\alpha_{\rm{EW}}$ 
channels are smaller by a factor
of four or so, compared to the tree-level $\alpha_{\rm{EW}}^2$
cross sections.  
Fig.~\ref{sigmabb_Tev} (bottom) shows the percentage
contributions of the   $\alpha_{\rm{S}}^3$, 
$\alpha_{\rm{EW}}^2$ and
$\alpha_{\rm{S}}^2\alpha_{\rm{EW}}$ subprocesses, with respect
to the leading $\alpha_{\rm{S}}^2$ ones, defined as the ratio of each of 
the former to the latter\footnote{In the case of
the $\alpha_{\rm{S}}^3$ corrections, we have used the two-loop 
expression for $\alpha_{\rm{S}}$ and a NLO fit for the
PDFs, as opposed to the one-loop formula and LO set for the 
other processes  (we adopted the GRV94 \cite{PDFs} 
PDFs with $\MSbar$ parameterisation).}.  
The $\alpha_{\rm{S}}^2\alpha_{\rm{EW}}$ 
terms represent a correction of the order
of the fraction of percent to the leading $\alpha_{\rm{S}}^2$ terms.
Clearly, at inclusive level, the effects of the Sudakov logarithms
are not large at Tevatron, 
this being mainly due to the fact that in the partonic
scattering processes the hard  scale involved is not much larger
than the $W$ and $Z$ masses.

Next, we study the forward-backward asymmetry, defined as
follows:
\begin{equation}\label{AFB}
A_{\rm{FB}}=
\frac{\sigma_+(p\bar p\to b\bar b)-\sigma_-(p\bar p\to b\bar b)}
     {\sigma_+(p\bar p\to b\bar b)+\sigma_-(p\bar p\to b\bar b)},
\end{equation}
where the subscript $+(-)$ iden\-ti\-fies events in which the $b$-jet
is produced with polar angle larger(smal\-ler) than 90 degrees respect to 
one of the two beam directions (hereafter, we use
the proton beam as positive $z$-axis).
The polar angle is defined in the 
CM frame of the hard partonic scattering. Notice that
we do not implement a jet algorithm, as we integrate over the entire phase
space available to the gluon. In practice, this corresponds to 
summing 
over the two- and three-jet contributions that one would extract from the
application of a jet definition. The solid curve in 
Fig.~\ref{sigmabb_AFB_Tev} (top) represents the sum of
 the tree-level contributions
only, that is, those of order $\alpha_{\rm{S}}^2$ and  
$\alpha_{\rm{EW}}^2$, whereas the dashed one also includes the
higher-order ones $\alpha_{\rm{S}}^3$ and
$\alpha_{\rm{S}}^2\alpha_{\rm{EW}}$. 

The effects of the one-loop weak corrections on this observable
are extremely large, 
as they are not only competitive with, if not larger than,
the tree-level weak contributions,
but also of opposite sign over most of the considered $p_T$ spectrum.
They are indeed comparable to the effects
through order $\alpha_{\rm{S}}^3$ \cite{Kuhn:1998kw}. 
In absolute terms, the asymmetry is of order $-4\%$ 
at the $W$, $Z$ resonance and
fractions of percent elsewhere, hence it should comfortably be measurable
after the end of Run 2.

Fig.~\ref{sigmabb_LHC} shows the same quantities as in 
Fig.~\ref{sigmabb_Tev}, now defined
at LHC energy. By a comparative reading, one may appreciate the 
following aspects. Firstly, the effects at LHC of the 
$\alpha_{\rm{S}}^2\alpha_{\rm{EW}}$ corrections are much larger
than the $\alpha_{\rm{EW}}^2$ ones already at inclusive level
 (see top of Fig.~\ref{sigmabb_LHC}), as
their absolute magnitude becomes of order $-2\%$ or so at large transverse
momentum (see bottom of Fig.~\ref{sigmabb_LHC}): 
clearly, logarithmic enhancements are at LHC much more 
effective than at Tevatron energy scales\footnote{Further notice at LHC
the dominance of the $gg$-induced one-loop terms, as compared
to the corresponding $q\bar q$-ones (top of Fig.~\ref{sigmabb_LHC}), 
contrary to the case of Tevatron, where they
were of similar strength (top of Fig.~\ref{sigmabb_Tev}).}. 

\section{Conclusions}
Altogether, the results presented here point to the relevance of one-loop
${\cal O}(\alpha_{\rm{W}})$ weak corrections for precision
analyses of three-jet rates at future high-luminosity LCs running at the
$Z$ pole, such as GigaZ,
of prompt-photon and neutral Drell-Yan events at both Tevatron and
LHC and of $b$-quark asymmetries (e.g., we have studied the forward-backward
one) at Tevatron.
(A suitable
definition of the $b$-quark forward-backward asymmetry -- see, e.g.,
Ref.~\cite{dittmar} -- may in fact reveal even larger effects
at the LHC.)

\end{document}